\documentclass[12pt]{article}

\usepackage{amsmath,amssymb}
\usepackage{mathrsfs}
\usepackage{amsfonts}
\usepackage{amsthm}
\usepackage{caption,subcaption}
\usepackage[colorlinks=true,
linkcolor=blue,
filecolor=black,      
urlcolor=blue,
citecolor=magenta]{hyperref}
\usepackage{tikz}
\usetikzlibrary{decorations.markings}
\usetikzlibrary{patterns}
\usepackage{color, soul}
\usepackage{feynmp-auto}
\usepackage{slashed}
\DeclareMathOperator{\arccosh}{arccosh}

\numberwithin{equation}{section}

\title{}
\author{}
\date{}
\begin{document}

\begin{titlepage}

\hfill\hbox{NORDITA 2021-038}\\
\vspace{0.1cm}
\hfill\hbox{UUITP-23/21}\\
\vspace{1.0cm}

\begin{center}
	{\Large \textbf{Infrared Divergences and\\the Eikonal Exponentiation}}
	\vskip 1.0cm 
	{\large Carlo Heissenberg$^{a,b}$} \\[0.7cm]
	
	{\it $^{a}$Nordita, Stockholm University and KTH Royal Institute of Technology,\\
		Hannes Alfv\'ens v\"ag 12,
		106 91 Stockholm,
		Sweden}\\
	\vspace{0.1cm}
	{\it $^{b}$Department of Physics and Astronomy, Uppsala University,\\ Box 516, 75120 Uppsala, Sweden}\\
	\vspace{0.1cm}
	e-mail: \textit{carlo.heissenberg@su.se}
\end{center}

\begin{abstract}
	The aim of this note is to explore the interplay between the eikonal resummation in impact-parameter space 
	and the exponentiation of infrared divergences in momentum space 
	for gravity amplitudes describing collisions of massive objects. The eikonal governs the classical dynamics relevant to the two-body problem, and its infrared properties are directly linked to the zero-frequency limit of the gravitational wave emission spectrum and to radiation-reaction effects. Combining eikonal and infrared exponentiations it is possible to derive these properties at a given loop order starting from lower-loop data. This is illustrated explicitly in $\mathcal{N}=8$ supergravity and in general relativity by deriving the divergent part of the two-loop eikonal from tree-level and one-loop elastic amplitudes.
\end{abstract}

\end{titlepage}


\section{Introduction}

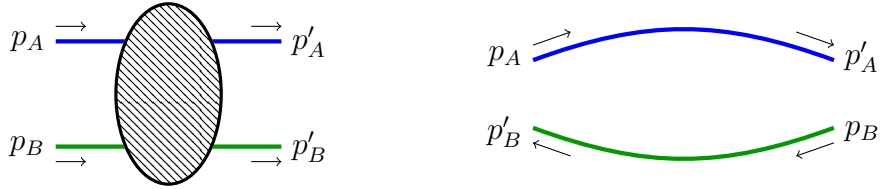
\begin{figure}
	\centering
	\begin{subfigure}{.45\textwidth}
	\centering
\begin{tikzpicture}
	\path [draw, ultra thick, blue] (-4.5,2.2)--(-1.5,2.2);
	\path [draw, ultra thick, green!60!black] (-4.5,.8)--(-1.5,.8);
	\filldraw[white, very thick] (-3,1.5) ellipse (.7 and 1.2);
	\filldraw[pattern=north west lines, very thick] (-3,1.5) ellipse (.7 and 1.2);
	\draw [->] (-4.5,2.4)--(-4.1,2.4);
	\draw [<-] (-1.5,2.4)--(-1.9,2.4);
	\draw [->] (-4.5,.6)--(-4.1,.6);
	\draw [<-] (-1.5,.6)--(-1.9,.6);
	\node at (-1.5,2.2)[right]{$p_A'$};
	\node at (-1.5,.8)[right]{$p_B'$};
	\node at (-4.5,2.2)[left]{$p_A$};
	\node at (-4.5,.8)[left]{$p_B$};
\end{tikzpicture}
	\end{subfigure}
\begin{subfigure}{.45\textwidth}
	\centering
\begin{tikzpicture}
	\draw[ultra thick, blue]  (-5,1.95).. controls (-3.5,2.5) and (-2.5,2.5) ..(-1,1.95);
	\draw[ultra thick, green!60!black] (-5,1.05).. controls (-3.5,.5) and (-2.5,.5) ..(-1,1.05);
	\draw [->] (-5,2.15)--(-4.5,2.325);
	\draw [<-] (-1,2.15)--(-1.5,2.325);
	\draw [<-] (-5,.85)--(-4.5,.675);
	\draw [->] (-1,.85)--(-1.5,.675);	
	\node at (-1,2)[right]{$p_A'$};
	\node at (-1,1)[right]{$p_B$};
	\node at (-5,2)[left]{$p_A$};
	\node at (-5,1)[left]{$p_B'$};
\end{tikzpicture}
\end{subfigure}
	\caption{\label{fig:kin} To the left, a diagrammatic picture of the elastic amplitude. To the right, a cartoon of two-body scattering in the center-of-mass frame. Thick colored lines represent massive particles.}
	\label{fig:scattering}
\end{figure}

Gravity amplitudes are constrained by two very different kinds of nonperturbative resummations that lead to the exponentiation of certain all-order contributions. The first is the eikonal exponentiation, which captures the classical limit of the scattering amplitude. For the collision of two objects with masses $m_A$, $m_B$ at a center-of-mass energy  $\sqrt s$, depicted in fig.~\ref{fig:scattering}, the classical regime is characterized by $G m^2_\ast \gg \hbar$, with $G$ the Newton constant and $m_\ast$ a scale of the order $m_{A,B}\lesssim m_\ast\lesssim \sqrt{s}$. This kinematic region lies outside the regime of validity of the standard ``small-$G$'' expansion, and the eikonal exponentiation provides a convenient tool to extract from it the rapidly oscillating phase factor $e^{2i\delta}$ characterizing the semiclassical limit of the amplitude. The perturbative nature of the calculations is recovered by looking at collisions with very large impact parameter $b\gg Gm_\ast$, which leads to a Post-Minkowski (PM) expansion $\delta = \delta_0 + \delta_1 + \delta_2+\cdots$ with $\delta_0\sim \frac{1}{\hbar}\,Gm_\ast^2$ plus higher-order PM corrections suppressed by additional powers of $\frac{Gm_\ast}{b}$. The systematic study of the eikonal resummation was initiated in the late eighties \cite{Amati:1987wq,tHooft:1987vrq,Amati:1987uf,Muzinich:1987in,Sundborg:1988tb,Amati:1990xe}, in the context of unitarity restoration in ultra-relativistic scattering, but it has been recently applied in amplitude-based approaches to the two-body problem and gravitational wave emission in general relativity (GR) and in its supersymmetric extensions \cite{Bjerrum-Bohr:2018xdl,KoemansCollado:2019ggb,Bern:2020gjj,Cristofoli:2020uzm,Bern:2020buy,Parra-Martinez:2020dzs,AccettulliHuber:2020oou,DiVecchia:2020ymx,Bern:2020uwk,DiVecchia:2021ndb,Bern:2021dqo,Kosmopoulos:2021zoq,DiVecchia:2021bdo,Bjerrum-Bohr:2021vuf}. 
Indeed, amplitude-based methods \cite{Goldberger:2004jt,Goldberger:2016iau,Damour:2017zjx,Luna:2017dtq,Cheung:2018wkq,Kosower:2018adc,Cristofoli:2019neg,Bjerrum-Bohr:2019kec,Mogull:2020sak,Huber:2020xny,Jakobsen:2021smu} have proved instrumental in advancing the state of the art in the PM analysis of the two-body problem \cite{Bern:2019nnu,Bern:2019crd,Bern:2021dqo}, leading to results that can be directly exported from scatterings to binary mergers, such as the conservative interaction Hamiltonian, or analytically continued from one case to the other \cite{Kalin:2019rwq,Kalin:2019inp,Herrmann:2021lqe,Herrmann:2021tct}.  

The second kind of exponentiation is that of infrared (IR) divergences that arise due to the long-range nature of gravitational interactions in four space-time dimensions. These infinities can be tamed introducing a suitable IR regulator such as a cutoff or a fictitious graviton mass, or moving slightly away from $D=4$. In this approach, IR divergences manifest themselves as poles in $\epsilon = \frac{1}{2}(4-D)$ although they could potentially mix with ultraviolet (UV) ones, associated to the usual short-distance singularities. 
However, infrared divergences can be neatly singled out  in momentum space by factoring out an exponential of the type $e^{\mathcal W}$ from the amplitude, where $\mathcal W$ is entirely determined by the one-loop IR divergences. The study of this exponentiation in GR dates back to Weinberg \cite{Weinberg:1965nx}, but a very similar phenomenon also takes place in supergravity theories \cite{Naculich:2008ew,Naculich:2011ry,White:2011yy,Akhoury:2011kq,Naculich:2013xa,Melville:2013qca,DiVecchia:2019myk,DiVecchia:2019kta}, whose amplitudes have recently attracted renewed interest due to their manifold formal and phenomenological applications. The maximally supersymmetric theory, $\mathcal N=8$, has been put forward as a potential UV finite theory of gravity in $D=4$ \cite{Bern:2006kd,Bern:2007hh,Bern:2009kd,Bern:2014sna,Bern:2018jmv}, and, especially in the context of the classical limit, it has proved to be a useful theoretical laboratory for developing new calculational tools and tackling conceptual issues in a technically simpler setup compared to GR \cite{Bern:2020gjj,Parra-Martinez:2020dzs,DiVecchia:2020ymx,DiVecchia:2021ndb,DiVecchia:2021bdo,Bjerrum-Bohr:2021vuf}.

Given these two a priori independent exponentiations, it is natural to wonder how they combine to provide constraints for  gravity amplitudes. This point was investigated in particular in \cite{DiVecchia:2019myk,DiVecchia:2019kta}, for massless $\mathcal N=8$ amplitudes, and led to the identification of all-loop structures together with a nontrivial interplay between the IR exponentiation and the eikonal to leading and subleading order.
The main goal of the present note is to initiate a similar study for massive $\mathcal N=8$ and GR amplitudes.
Besides being more directly related to the scattering of compact objects, and thus to the problem of gravitational-wave emission by ``bremsstrahlung'', amplitudes with massive states bring along an interesting novelty since they open up a new kinematic region $m_{A,B}\lesssim m_\ast \lesssim \sqrt s$, compared to the massless case where effectively $m_\ast=\sqrt s$. 

As a first step towards this goal, it is natural to apply the exponentiation of IR divergences to the calculation of the divergent part of the 3PM eikonal $\delta_2$, recently obtained in ref.~\cite{DiVecchia:2021ndb} using the method of unitarity cuts and Weinberg's soft limit.
As discussed in that reference, the divergent part of $\operatorname{Im}\delta_2$ is directly linked to the total number of emitted gravitons (plus additional massless particles in the supersymmetric setup) and to the zero-frequency limit of the energy emission spectrum. Moreover, it determines via analyticity and crossing symmetry the dissipative radiation-reaction corrections to the 3PM deflection angle  \cite{Damour:2020tta,DiVecchia:2021ndb,DiVecchia:2021bdo,Herrmann:2021tct}.
The explicit expressions for $\delta_2$ in $\mathcal N=8$ supergravity (taking orthogonal Kaluza--Klein momenta in the toroidal compactification as in Section~\ref{sec:sugra}) and in GR read
\begin{align}
	\left(2\delta_2\right)_{\mathcal N=8}
	&= 
	-\frac{i}{\pi\epsilon}\,\frac{G {\beta}_{\mathcal N=8}^2}{b^2(\sigma^2-1)^2} 
	\left[ \sigma^2 +\sigma (\sigma^2-2)\, \frac{\arccosh\sigma}{\sqrt{\sigma^2-1}} \right],
	\label{3.6}
\\
	(2 \delta_2)_\mathrm{GR}
	&= 
	-\frac{i}{\pi \epsilon }\, \frac{G {\beta}_\mathrm{GR}^2}{2b^2(\sigma^2-1)^2}
	\left[ \frac{ 8 -5 \sigma^2}{3}  +
	\sigma(2\sigma^2-3)\, \frac{\arccosh\sigma}{\sqrt{\sigma^2-1}} \right],
	\label{3.2GR}
\end{align} 
up to $\mathcal{O}(\epsilon^0)$, where (see fig.~\ref{fig:scattering})
\begin{equation}\label{}
	\sigma= -\frac{p_Ap_B}{m_Am_B}\,,\quad
	\beta_{\mathcal N=8} = 4 G m_A m_B \sigma^2\,,\quad
	\beta_\mathrm{GR} = 4  G m_A m_B \left(\sigma^2-\tfrac12\right)\,.
\end{equation}
The main result of this note is to show how both these expressions can be derived combining eikonal and IR exponentiations, using only some information about the tree-level and one-loop elastic amplitudes. In particular, this
strategy will show that they are indeed purely imaginary as expected on physical grounds.

The approach discussed here is thus complementary to the one employed in \cite{DiVecchia:2021ndb}. There, the basic idea was to focus directly on the imaginary part of the eikonal, which can be obtained from the three-particle cut involving appropriate ``squares'' of five-point amplitudes. Moreover, in order to isolate the divergent contribution, it was sufficient to restrict to the Weinberg soft limit.
In fact, going beyond the Weinberg limit, the same method based on the three-particle cut was employed to evaluate the full $\operatorname{Im}\delta_2$ including $\mathcal O(\epsilon^0)$ terms both in $\mathcal N=8$ and in GR \cite{DiVecchia:2021bdo}.
On the other hand, the present approach
only relies on exponentiations of the elastic amplitude without resorting to unitarity cuts involving higher-point amplitudes. Moreover, unlike the one based on unitarity, this method treats real and imaginary contributions in a completely democratic way and thus provides a check that the divergent parts \eqref{3.6} and \eqref{3.2GR} are indeed purely imaginary.
In fact, while exchanges of real soft quanta are tailored to the evaluation of imaginary IR divergences, the exponentiation of the complete set of one-loop IR poles allows one to single out both real and imaginary divergences.

Both methods allow one to bypass much more involved loop calculations by selectively focusing on the evaluation of specific pieces of the amplitudes. Both can be extended to higher-loop orders, and can serve as complementary methods for constraining the IR behavior of the amplitude and of the eikonal. A special motivation for generalizing these type of analyses to three-loop order is the apparent clash between the two exponentiations that has been pointed out for massless amplitudes in ref.~\cite{DiVecchia:2019kta}, which ought to manifest itself for all loop orders $L\ge3$. Methods and ideas presented here will hopefully clarify whether this issue is special to the massless setup or if it is shared by the massive setup.

The paper is organized as follows.
Section~\ref{sec:eikonal} is devoted to a brief pedagogical summary of the eikonal exponentiation.  Eqs.~\eqref{3.6} and \eqref{3.2GR} are derived combining the eikonal and the IR resummations in Section~\ref{sec:resum}. Notations and conventions are collected in Appendix~\ref{app:conventions}, while Appendix~\ref{app:IR} contains the explicit evaluation of some useful IR-divergent integrals.

\section{Eikonal Exponentiation}
\label{sec:eikonal}

The spinless $2\to2$ amplitude
$\mathcal A(s,t)$ in the usual momentum space representation
is a function of the masses and of the Mandelstam invariants $s$ and $t$, where
\begin{equation}\label{}
	s=-(p_A+p_B)^2=m_A^2+2m_Am_B\sigma+m_B^2\,, \qquad t=-q^2\,,\qquad q^\mu = p_A'^\mu - p_A^\mu\,.
\end{equation}
In the eikonal approach, one considers the amplitude in impact-parameter space, defining a transverse
$(2-2\epsilon)$-dimensional Fourier-transform (see Appendix~\ref{app:conventions})
\begin{equation}\label{2dFT}
	\tilde{\mathcal A}(s, b)=\frac{1}{4\Omega}  \int e^{ibq_\perp} \mathcal A\left(s,-q_\perp^2\right) \, \hat dq_\perp\,,
	\qquad
	\Omega=  m_A m_B \sqrt{\sigma^2-1}\,.
\end{equation}
The classical regime requires that the effective quantum wavelength of each colliding body be much smaller than its size. The latter in turn ought to be negligible compared to the relative separation between the two bodies, sized by the impact parameter $b$, in order to ensure weak gravitational interactions. Reinstating momentarily $\hbar$, these two conditions identify the hierarchy of length scales
\begin{equation}\label{hierarchy}
	\frac{\hbar}{m_{\ast}}\ll 
	G m_{\ast} b^{2\epsilon} \ll b
\end{equation}
where $m_\ast$ is an energy scale that can be taken as $m_{A,B}\lesssim m_\ast \lesssim \sqrt s$, so that $\frac\hbar{m_\ast}$ can be regarded as the Compton wavelength and $Gm_\ast b^{2\epsilon}$ as the effective Schwarzschild radius.
The large dimensionless quantity characterizing the classical limit is thus $\frac{1}{\hbar}G m_\ast^2 b^{2\epsilon}$, while the small parameter sizing PM corrections is $G m_\ast b^{2\epsilon-1}$. 

The eikonal resummation postulates that, in the regime discussed above, the amplitude in impact-parameter space exponentiates according to
\begin{equation}\label{eikonalexponentiation}
	1+i\tilde{\mathcal A}(s,b)=e^{2i\delta(s,b)}\big(
	1+2i\Delta(s,b)
	\big),
\end{equation}
were $\delta(s,b)$ is a classical quantity that scales like $\frac{1}{\hbar}G m_\ast^2 b^{2\epsilon}$ to leading order in $G$ and only acquires PM corrections to higher orders,
\begin{equation}\label{deltaL}
	\delta=\delta_0+\delta_1+\delta_2+\cdots\,,\qquad
	\delta_{L} \sim \frac{1}{\hbar}\, G m_\ast^2b^{2\epsilon}\left(\frac{G m_\ast}{b^{1-2\epsilon}}\right)^{L},
\end{equation}
while $\Delta$ is a quantum remainder, 
\begin{equation}\label{}
	\Delta=\Delta_1+\Delta_2+\cdots\,,\qquad
	\Delta_{L} \sim \sum_{k>0} \left(\frac{1}{\hbar}\,G m_\ast^2  b^{2\epsilon}\right)^{1-k}\left(\frac{G m_\ast}{b^{1-2\epsilon}}\right)^{L+k}.
\end{equation}

Matching the formal small-$G$ expansion of
eq.~\eqref{eikonalexponentiation} with the standard loop expansion of the amplitude 
\begin{equation}\label{}
	\mathcal A = 	\mathcal A_0+	\mathcal A_1+	\mathcal A_2+\cdots
\end{equation}
the eikonal exponentiation amounts to the following relations, up to two-loop order, 
\begin{align}\label{eik0}
	i\tilde{\mathcal A}_0 &=2i\delta_0\,,\\
	\label{eik1}
	i\tilde{\mathcal A}_1 &= \frac{(2i\delta_0)^2}{2!}
	+2i\delta_1 + 2i\Delta_1\,,\\
	\label{eik2}
	i\tilde{\mathcal A}_2 &= \frac{(2i\delta_0)^3}{3!}
	+ 2i\delta_0 \,2i\delta_1 + \big[2i\delta_2 + 2i\delta_0\, 2i\Delta_1\big] + 2i\Delta_2\,.
\end{align}
Terms of the type $(2i\delta_0)^2$ or $2i\delta_0\,2i\delta_1$ arise from the expansion of $e^{2i\delta}$ and are usually referred to as ``superclassical'' because they are even more singular in the formal limit $\hbar\to0$ compared to $\delta$ itself.
The exponentiation of the leading superclassical term $i\tilde{\mathcal A}_{n-1} \sim \frac{1}{n!}\,(2i\delta_0)^n+\cdots$ has been proved on general grounds \cite{Levy:1969cr,Kabat:1992tb,Akhoury:2013yua,Bjerrum-Bohr:2018xdl,Fazio:2021aai}, while the subleading terms have been checked in a variety of different setups up to three loops \cite{Amati:1990xe,Collado:2018isu,DiVecchia:2019myk,DiVecchia:2019kta,Bern:2020gjj,Bern:2021dqo,DiVecchia:2021bdo}, supporting the strong belief that eq.~\eqref{eikonalexponentiation} indeed captures the right classical asymptotics of the amplitude.

In the limit \eqref{hierarchy}, one effectively focuses on the large-$b$ expansion of $\tilde{\mathcal A}$ or, equivalently, on the small-$q$ expansion of $\mathcal A$. Moreover, via Fourier transform, only non-analytic terms in $q^2$ give rise to long-range tails in $b$ that are relevant to macroscopic classical interactions, and therefore terms proportional to $(q^2)^n$ for non-negative integer $n$ can be safely dropped.

In view of eq.~\eqref{eik2}, one can conveniently split $\tilde{\mathcal A}_2$ as follows
\begin{equation}\label{}
	\tilde{\mathcal A}_2
	=
	\big(\tilde{\mathcal A}_2\big)_\textrm{sscl.}+
	\big(\tilde{\mathcal A}_2\big)_\textrm{scl.}+
	\big(\tilde{\mathcal A}_2\big)_\textrm{cl.}+\cdots\,,
\end{equation} 
isolating classical and (super)-superclassical terms according to
\begin{equation}\label{}
	i\big(\tilde{\mathcal A}_2\big)_\textrm{sscl.}= \frac{(2i\delta_0)^3}{3!}\,,
	\quad
	i\big(\tilde{\mathcal A}_2\big)_\textrm{scl.}
	= 2i\delta_0 \,2i\delta_1\,,
	\quad
	i\big(\tilde{\mathcal A}_2\big)_\textrm{cl.}
	=
	2i\delta_2 + 2i\delta_0\, 2i\Delta_1
\end{equation}
and neglecting quantum terms.
Concretely, in order to retrieve $2i\delta_2$, it is sufficient to evaluate the $\mathcal O(b^{-2(1-3\epsilon)})$ terms in $\tilde{\mathcal A}_2$, which arise from those of order $\mathcal O((q^2)^{-2\epsilon})$ in $i\mathcal A_2$, and subtract from them the contribution $2i\delta_0\,2i\Delta_1$ obtained from lower loop orders, focusing on the $\mathcal O(b^{-2(1-2\epsilon)})$ leading terms of $2i\Delta_1$.

\section{Exponentiation of IR Divergences}
\label{sec:resum}
In this section the exponentiation of IR divergences is employed to obtain the expressions displayed above for $2\delta_2$ as $\epsilon\to0$, eq.~\eqref{3.6} for $\mathcal N=8$ with orthogonal Kaluza--Klein momenta and eq.~\eqref{3.2GR} for GR. 
As briefly mentioned in the introduction and explained in detail in \cite{DiVecchia:2021ndb,DiVecchia:2021bdo}, the IR-divergent part of $\delta_2$ directly determines the number of emitted massless quanta and the zero-frequency limit of the energy emission spectrum to 3PM. Furthermore, via analyticity and crossing-symmetry constraints, it allows one to derive the 3PM radiation-reaction effects appearing in the deflection angle. These are corrections that must be included because the colliding particles not only deviate from straight lines due to their ``potential'' interactions, but also emit radiation and thus decelerate.

The technical key-point of the present analysis is that, for the $2\to2$ amplitude, the two-loop contributions relevant to the divergent part of $2\delta_2$ are those proportional to $\frac{1}{\epsilon}\,\log q^2$  in the small-$q$, small-$\epsilon$ expansion and they can be deduced from certain IR divergent or non-analytic lower-order terms, as illustrated below.

\subsection{Maximal Supergravity}
\label{sec:sugra}
Massive $\mathcal N=8$ amplitudes can be conveniently obtained by dimensional reduction of the corresponding massless type II amplitudes \cite{Caron-Huot:2018ape,Parra-Martinez:2020dzs}.
Starting from ten-dimensional type II supergravity, one scatters Kaluza--Klein (KK) scalars whose $(10-2\epsilon)$-dimensional momenta read as follows:
\begin{equation}
	\label{kin}
	P_A= (p_A;0,0,0,0,0,m_A)\,,\qquad P_B = (p_B;0,0,0,0,m_B ,0)\;,
\end{equation}
where the last six entries refer to the compact directions and $m_A$, $m_B$ correspond to the first two excited KK states. 
Since
\begin{equation}\label{}
	P_A^2 = p_A^2+m_A^2=0\,,\qquad 
	P_B^2 = p_B^2+m_B^2=0\,,
\end{equation}
this provides $p_A$, $p_B$ with the desired effective masses $m_A$, $m_B$ in $4-2\epsilon$ dimensions.
 
In order to deduce the infrared properties of the two-loop amplitude, we shall exploit a momentum-space exponentiation akin to the one discussed in \cite{DiVecchia:2019myk,DiVecchia:2019kta} in the context of the massless theory, according to which the elastic amplitude $\mathcal A=\mathcal A_0+\mathcal A_1+\mathcal A_2+\cdots$ in momentum space can be cast in the form 
\begin{equation}\label{N8exp}
	\mathcal A = \mathcal A_0\, e^{\hat {\mathcal A}_1} \left(1+\hat {\mathcal A}_2+\cdots\right),
\end{equation}
where the exponent
$\hat {\mathcal A}_1=\mathcal A_1/\mathcal A_0$ contains all infrared divergences and $\hat {\mathcal A}_2$ denotes an IR-finite two-loop contribution. 

The tree-level amplitude for the $s$-$u$ symmetric scattering of an axion and a dilaton reads \cite{DiVecchia:2021bdo}
\begin{equation}\label{}
	\mathcal A_0
	=
	8\pi G
	\,\frac{(s-m_A^2-m_B^2)^4+(u-m_A^2-m_B^2)^4-t^4}{2(s-m_A^2-m_B^2)\,t\,(u-m_A^2-m_B^2)}
\end{equation}
and in the near-forward limit  $q\to0$ this reduces to
\begin{equation}\label{A0simple}
	\mathcal A_0 = \frac{16 \pi G m_A m_B \sigma }{q^2}\,(2 m_A m_B \sigma -q^2)+\mathcal O(q^2)\,,
\end{equation} 
where, together with the non-analytic $\frac1{q^2}$ term that is needed to determine $2\delta_0$, we retain the first analytic term as well.

The one-loop amplitude can be represented as follows
\begin{equation}\label{}
	\begin{split}
\mathcal A_1 &= \frac12 \,\frac{(8\pi G)^2}{(4\pi)^{2-\epsilon}} \left[
(s-m_A^2-m_B^2)^4+(u-m_A^2-m_B^2)^4-t^4
\right]\\
&\times \left(\,
\begin{gathered}
\begin{tikzpicture}
	\path [draw, ultra thick] (-.65,0)--(.65,0);
	\path [draw, ultra thick] (-.65,1)--(.65,1);
	\path [draw, ultra thick] (.5,0)--(.5,1);
	\path [draw, ultra thick] (-.5,1)--(-.5,0);
\end{tikzpicture}
\end{gathered}
+
\begin{gathered}
	\begin{tikzpicture}
		\path [draw, ultra thick] (-.65,0)--(.65,0);
		\path [draw, ultra thick] (-.65,1)--(.65,1);
		\path [draw, ultra thick] (.5,1)--(-.5,0);
		\path [draw, ultra thick] (-.5,1)--(.5,0);
	\end{tikzpicture}
\end{gathered}
+
\begin{gathered}
	\begin{tikzpicture}
		\path [draw, ultra thick] (-.65,1)--(-.5,1)--(.5,0)--(.65,0);
		\path [draw, ultra thick] (-.65,0)--(-.5,0)--(.5,1)--(.65,1);
		\path [draw, ultra thick] (-.5,1)--(-.5,0);
		\path [draw, ultra thick] (.5,1)--(.5,0);
	\end{tikzpicture}
\end{gathered}
\,
\right),
	\end{split}
\end{equation}
where the thick-line diagrams are scalar one-loop topologies with numerator equal to one, where one should integrate over the $(4-2\epsilon)$-dimensional loop momentum and sum over the possible ways of assigning the KK momenta, compatibly with momentum conservation. However, for the present purposes, these sums can be restricted to the lowest-lying excitations compatible with the external states \eqref{kin}, with the last two momentum entries equal to
\begin{equation}\label{}
	(0,0)\,,\qquad (\pm m_B,0)\,,\qquad
	(0,\pm m_A)\,,\qquad
	(\pm m_B, \pm m_A)\,.
\end{equation}
As discussed in \cite{Parra-Martinez:2020dzs}, this truncation to the lowest-lying massive excitations is not expected to provide a fully consistent quantum theory, since there is no parametric separation between such states and the full tower of KK excitations. However, it is expected to preserve the properties of the eikonal capturing the classical dynamics. 

In view of eq.~\eqref{N8exp}, the two-loop IR-divergent part of $\mathcal A_2$ is fully captured by  
\begin{equation}\label{A2A1A0}
	\mathcal A_2 \simeq \frac{1}{2}\mathcal A_0\, \hat {\mathcal A}_1^2 = \frac{{\mathcal A}_1^2}{2\mathcal A_0} \,,
\end{equation}
and in order to retrieve its $\frac{1}{\epsilon}\,\log q^2$ terms we may disregard terms in the one-loop amplitude that are analytic in $q^2$ or IR finite.
Non-analytic or IR divergent terms can only arise when at least one massless line is present in the loop.
Assigning the KK momenta compatibly with this requirement leads to the topologies illustrated in fig.~\ref{fig:diagrams}, together with their crossing symmetry counterparts that can be obtained interchanging, say, $p_A$ with $-p_A'$ (i.e.~interchanging the two endpoints of the blue line). These one-loop topologies can be constructed out of the tree-level building blocks in fig.~\ref{fig:buildingblocks}.
We note for later convenience that since $p_B'=p_B-q$ and $2p_Bq=q^2$ in view of the mass-shell constraints $p_B'^2=-m_B^2=p_B^2$, crossing symmetry $p_A \to -p_A'=-p_A-q$ is equivalent to
\begin{equation}\label{crossingsigma}
	\sigma \to -\sigma-i0 + \frac{q^2}{2m_A m_B}\,,
\end{equation}
where the $-i0$ is included in order to resolve the branch cut discontinuity across $\sigma>1$, in particular
\begin{equation}\label{crossingarccosh}
	\arccosh\sigma \to i\pi-\arccosh\sigma + \mathcal O(q^2)\,.
\end{equation}
In the following, crossing symmetry transformations will be performed by either $p_A\to-p_A'$ or, equivalently, by $p_B\to-p_B'$, depending on convenience.
\begin{figure}
	\centering
	\begin{subfigure}[b]{.18\textwidth}
		\centering
		\begin{tikzpicture}
			\path [draw, ultra thick, green!60!black] (-.65,0)--(.65,0);
			\path [draw, ultra thick, blue] (-.65,1)--(.65,1);
			\path [draw,thin] (.5,.97)--(.5,.03);
			\path [draw,thin] (-.5,.97)--(-.5,.03);
		\end{tikzpicture}
		\caption{}
		\label{fig:box}
	\end{subfigure}
	\begin{subfigure}[b]{.18\textwidth}
		\centering
		\begin{tikzpicture}
			\path [draw, ultra thick, blue] (-.65,1)--(-.5,1)--(.5,.04)--(.5,1)--(.65,1);
			\path [draw, ultra thick, green!60!black] (-.65,0)--(-.5,0)--(.45,.96)--(.45,0)--(.65,0);
			\path [draw,thin] (-.5,.97)--(-.5,.03);
		\end{tikzpicture}
		\caption{}
		\label{fig:rightpapillon}
	\end{subfigure}
	\begin{subfigure}[b]{.18\textwidth}
		\centering
		\begin{tikzpicture}
			\path [draw, ultra thick, blue] (.65,1)--(.5,1)--(-.5,.04)--(-.5,1)--(-.65,1);
			\path [draw, ultra thick, green!60!black] (.65,0)--(.5,0)--(-.45,.96)--(-.45,0)--(-.65,0);
			\path [draw,thin] (.5,.97)--(.5,.03);
		\end{tikzpicture}
		\caption{}
		\label{fig:leftpapillon}
	\end{subfigure}
	\begin{subfigure}[b]{.18\textwidth}
		\centering
		\begin{tikzpicture}
			\path [draw, ultra thick, blue] (-.65,1)--(-.5,1)--(-.5,.04)--(.5,.04)--(.5,1)--(.65,1);
			\path [draw, ultra thick, green!60!black] (.65,0)--(-.65,0);
			\path [draw,thin] (.47,1)--(-.47,1);
		\end{tikzpicture}
		\caption{}
		\label{fig:lowerbox}
	\end{subfigure}
	\begin{subfigure}[b]{.18\textwidth}
		\centering
		\begin{tikzpicture}
			\path [draw, ultra thick, green!60!black] (-.65,0)--(-.5,0)--(-.5,.96)--(.5,.96)--(.5,0)--(.65,0);
			\path [draw, ultra thick, blue] (.65,1)--(-.65,1);
			\path [draw,thin] (.47,0)--(-.47,0);
		\end{tikzpicture}
		\caption{}
		\label{fig:upperbox}
	\end{subfigure}
	\caption{Topologies relevant for the calculation of the IR divergences of $\mathcal{A}_1$ in massive $\mathcal N=8$ supergravity. External states are as in fig.~\ref{fig:kin}. The squared mass is zero for the thin black lines, $m_A^2$ ($m_B^2$) for the thick blue (green) lines, and $m_A^2+m_B^2$ for the mixed lines.	
	}
	\label{fig:diagrams} 
\end{figure}
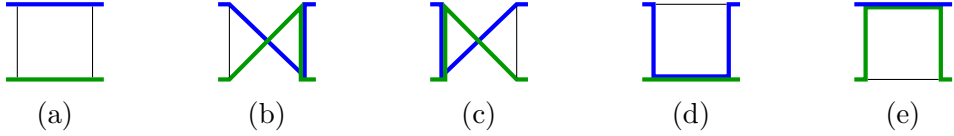
\begin{figure}
	\centering
	\begin{subfigure}[b]{.18\textwidth}
		\centering
		\begin{tikzpicture}
			\path [draw, ultra thick, green!60!black] (-.3,0)--(.3,0);
			\path [draw, ultra thick, blue] (-.3,1)--(.3,1);
			\path [draw,thin] (0,0)--(0,1);
		\end{tikzpicture}
	\end{subfigure}
	\begin{subfigure}[b]{.18\textwidth}
		\centering
		\begin{tikzpicture}
			\path [draw, ultra thick, green!60!black] (-.3,0)--(-.03,0)--(-.03,1)--(-.3,1);
			\path [draw, ultra thick, blue] (.3,0)--(.03,0)--(.03,1)--(.3,1);
		\end{tikzpicture}
	\end{subfigure}
	\begin{subfigure}[b]{.18\textwidth}
		\centering
		\begin{tikzpicture}
			\path [draw, ultra thick, green!60!black] (-.3,0)--(-.03,0)--(-.03,1)--(.3,1);
			\path [draw, ultra thick, blue] (.3,0)--(.03,0)--(.03,1)--(-.3,1);
		\end{tikzpicture}
	\end{subfigure}
	\caption{Building blocks for the topologies in fig.~\ref{fig:diagrams} and their crossed counterparts.
	}
	\label{fig:buildingblocks} 
\end{figure}
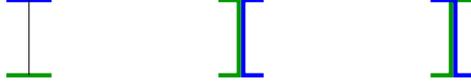

The box diagram corresponding to fig.~\ref{fig:box} reads 
\begin{equation}\label{completebox}
	I_{\mathrm{(a)}}
	=
	\int_\ell \frac{1}{(\ell^2+2 p_A\ell-i0)(\ell^2-2 p_B\ell-i0)(\ell^2-i0)((\ell-q)^2-i0)}\,,
\end{equation} 
where $\int_\ell$ is defined in eq.~\eqref{loopintegral}. The Feynman $-i0$ prescription explicitly spelled out in eq.~\eqref{completebox} will often be left implicit from now on.
The integral $I_{\mathrm{(a)}}$, together with the crossed box $I_{\overline{\mathrm{(a)}}}$,  was discussed in the small-$q$ expansion in \cite{Collado:2018isu,Cristofoli:2020uzm,DiVecchia:2021bdo}, using the method of regions \cite{Beneke:1997zp,Smirnov:2002pj}. Calculations in the context of the classical limit typically focus on the soft region, $\ell \sim \mathcal O(q)\ll m_\ast$, which captures non-analytic terms in the small-$q^2$ expansion and yields
\begin{equation}\label{IaIbara}
	I_{\mathrm{(a)}}^s
	=
	\frac{c_\mathrm{II}}{\epsilon(q^2)^{1+\epsilon}}
	 + \cdots\,,
	 	\qquad
	 I_{\overline{\mathrm{(a)}}}^s
	 	=
	 	\frac{1}{\epsilon(q^2)^{\epsilon}}\left(\frac{c_{\mathrm{X}}}{q^2}+c_0\right)+\cdots\,,
	\end{equation}
with
	\begin{equation}\label{cIIcX}
		c_{\mathrm{II}}=\frac{\operatorname{arccosh}\sigma-i\pi}{m_A m_B\sqrt{\sigma^2-1}}
		\,,\quad
		c_{\mathrm{X}}=\frac{-\operatorname{arccosh}\sigma}{m_A m_B\sqrt{\sigma^2-1}}\,,
		\quad
		c_{0}=\frac{-\sigma \operatorname{arccosh}\sigma+\sqrt{\sigma^2-1}}{2m_A^2m_B^2(\sigma^2-1)^{\frac{3}{2}}}\,.
	\end{equation}
The dots in \eqref{IaIbara} stand for terms $\mathcal O\left(q^{1-2\epsilon}\right)$ or smaller and for higher-order terms as $\epsilon\to0$.
The hard region $\ell \sim \mathcal O(m_\ast)\gg q$ only yields analytic terms, but can in principle produce IR-divergent terms relevant to the present analysis. In this respect, it is sufficient to focus on the $\mathcal O(q^0)$ term arising from the leading hard-region contribution to \eqref{completebox}, or equivalently its crossed counterpart $I_{\overline{\mathrm{(a)}}}$,
\begin{equation}\label{}
	I_{\overline{\mathrm{(a)}}}^h=\int_\ell \frac{1}{(\ell^2+2 p_A\ell)(\ell^2+2 p'_B\ell)(\ell^2)^2}\,,
\end{equation}
where to leading order one can approximate $p'_B\simeq p_B$.
Using Schwinger parameters this integral reduces to \begin{equation}\label{}
	I_{\overline{\mathrm{(a)}}}^h
	=
	\Gamma(2+\epsilon)
	\int_{\mathbb R_+^2}\frac{(1+x_A+x_B)^{2\epsilon}}{(m_A^2x_A^2+2m_Am_B\sigma x_A x_B+m_B^2x_B^2)^{2+\epsilon}}\,dx_A\,dx_B\,,
\end{equation}
and letting $x_A = \frac{\lambda\, x}{1+x}$, $x_B = \frac{\lambda }{1+x}$ this can be cast in the form
\begin{equation}\label{}
I_{\overline{\mathrm{(a)}}}^h
	=
	\frac{\Gamma(2+\epsilon)}{(1+2\epsilon)(2+2\epsilon)}
	\int_0^\infty
	\frac{(1+x)^{2+2\epsilon}\,dx}{(m_A^2x^2+2 m_A m_B \sigma x + m_B^2)^{2+\epsilon}}\,,
\end{equation}
which is manifestly finite for $\epsilon=0$ (see also Appendix~B of \cite{Cristofoli:2020uzm}). Therefore, hard-region contributions to $I_\mathrm{(a)}$ and $I_{\overline{\mathrm{(a)}}}$ can be  safely dropped.

The remaining topologies in fig.~\ref{fig:diagrams} give rise to $q$-independent IR divergent contributions that can be estimated by means of a small upper cutoff $\Lambda$. Lef us start from the crossed analog of \ref{fig:rightpapillon}, 
\begin{equation}\label{papib}
	I_{\overline{\mathrm{(b)}}}
	=
	\int_\ell
	\frac{1}{(\ell^2+2p_A\ell)(\ell^2+2p'_B\ell)\ell^2
		[\ell^2+2(p_A+p_B)\ell-2m_A m_B \sigma]}\,,
\end{equation}
where we can approximate $p_B'\simeq p_B$ up to $\mathcal O(q)$.
Infrared divergent contributions to the integral over $\ell=(\ell^0,\vec \ell\,)$ can only arise from the small-$|\vec\ell\,|$ region. To estimate them we can therefore place a small cutoff on $|\vec\ell\,|$ and retain only the leading order in $\ell$ in the denominator of \eqref{papib}, obtaining
\begin{equation}\label{}
	I_{\overline{\mathrm{(b)}}}^\mathrm{IR}=
	\frac{i}{2m_A m_B \sigma\pi^2}
	\int_{\Lambda}\frac{d\ell}{(2p_A\ell-i0)(2p_B\ell-i0)(\ell^2-i0)}\,,
\end{equation}
with $\int_\Lambda d\ell$ as in \eqref{loopintegral}.
Using the general expression \eqref{general} for $p_A=p$ and $p_B=p'$, one obtains
\begin{equation}\label{Iright}
	I_{\overline{\mathrm{(b)}}}^\mathrm{IR}=
	\frac{1}{\epsilon}
	\frac{\arccosh\sigma}{4m_A^2m_B^2\sigma \sqrt{\sigma^2-1}}\,,
\end{equation}
to leading order in $\epsilon$ (note that the dependence on the upper cutoff $\Lambda$ has dropped out)
and a similar contribution comes from the diagram displayed in fig.~\ref{fig:rightpapillon} obtained using the crossing-symmetry equations \eqref{crossingsigma}, \eqref{crossingarccosh} in \eqref{Iright}
\begin{equation}\label{Irightbar}
		I_{\overline{\mathrm{(b)}}}^\mathrm{IR}=
	\frac{1}{\epsilon}
	\frac{\arccosh\sigma-i\pi}{4m_A^2m_B^2\sigma \sqrt{\sigma^2-1}}\,.
\end{equation}
Adding \eqref{Iright} and \eqref{Irightbar}, and multiplying by two due to the identical contributions arising from fig.~\ref{fig:leftpapillon} leads to
\begin{equation}\label{}
		I_{\mathrm{(b)}}^\mathrm{IR}+	I_{\overline{\mathrm{(b)}}}^\mathrm{IR}+	I_{\mathrm{(c)}}^\mathrm{IR}+	I_{\overline{\mathrm{(c)}}}^\mathrm{IR}
		=
		\frac{1}{\epsilon}\,
		\frac{2\operatorname{arccosh}\sigma-i\pi}{2 m_A^2 m_B^2\sigma\sqrt{\sigma^2-1}}
		\equiv\frac{d}{\epsilon}\,.
\end{equation}

Finally, to discuss the contributions arising from \ref{fig:lowerbox} and \ref{fig:upperbox},
let us start from
\begin{equation}\label{}
	I_{\overline{\mathrm{(d)}}}
	=
	\int_\ell \frac{1}{\ell^2(\ell^2+2p_A\ell)(\ell^2+2p_A'\ell)[\ell^2+2(p_A-p_B')\ell-2p_Ap_B']}\,,
\end{equation}
where one can safely approximate $p_A'\simeq p_A$, $p_B'\simeq p_B$ and $-p_Ap_B'\simeq m_A m_B \sigma$ to leading order in $q$. Following the same strategy as above in order to highlight the IR divergent part,
\begin{equation}\label{}
	I_{\overline{\mathrm{(d)}}}^\mathrm{IR}
	=
	-
	\frac{i}{2m_A m_B \sigma\pi^2}
	\int_{\Lambda}\frac{d\ell}{(2p_A\ell-i0)^2(\ell^2-i0)}\,,
\end{equation}
where eq.~\eqref{special} for $p_A=p$ then yields
\begin{equation}\label{}
	I_{\overline{\mathrm{(d)}}}^\mathrm{IR}
	=
	-\frac{1}{\epsilon}
	\frac{1}{4m_A^3 m_B \sigma}\,.
\end{equation}
This contribution clearly vanishes after summing its crossing-symmetry counterpart $I_{\mathrm{(d)}}^\mathrm{IR}$, due to \eqref{crossingsigma}.
The divergent part arising from $I_{\mathrm{(e)}}$, $I_{\overline{\mathrm{(e)}}}$ drops out in a similar fashion.

All in all, for small $q$ and as $\epsilon\to0$,
\begin{equation}\label{A1simple}
	\mathcal A_1 = (4 m_A m_B \sigma)^3 G^2 (m_A m_B \sigma-q^2) \left[\frac{1}{\epsilon(q^2)^{\epsilon}}\left(\frac{c_{2}}{q^2}+c_0\right)+\frac{d}{\epsilon}+\cdots\right],
\end{equation} 
where $c_2 = c_\mathrm{II}+c_\mathrm{X}$.
The leading divergent, non analytic terms of the two-loop amplitude are then obtained via eq.~\eqref{A2A1A0},
\begin{equation}\label{}
	\mathcal A_2 
	\simeq
	\frac{(2m_Am_B\sigma)^6G^3}{\pi\epsilon^2}
	\left[
	\frac{(c_2)^2}{(q^2)^{1+2\epsilon}}
	+
	\frac{2 c_2}{(q^2)^{2\epsilon}}
	\left(
	c_0-\frac{3 c_2}{4m_A m_B\sigma}
	\right)
	+
	\frac{2c_2\,d}{(q^2)^{\epsilon}}
	\right]
\end{equation}
In particular, this captures the $\frac{1}{\epsilon}\log q^2$ terms in $\mathcal A_2$. Going to impact parameter space using \eqref{2dFT} and \eqref{basic} then yields
\begin{equation}\label{}
	\tilde{\mathcal A}_2
	\simeq
	\big(\tilde{\mathcal A}_2\big)_\mathrm{sscl.}
	+
	\big(\tilde{\mathcal A}_2\big)_\mathrm{cl.}\,,
\end{equation}
where, up to leading order in $\epsilon$, 
\begin{equation}\label{}
	\big(\tilde{\mathcal A}_2\big)_\mathrm{sscl.}
	=
	-
	\frac{4(m_Am_B\sigma)^5 \sigma G^3}{3 \sqrt{\sigma^2-1}\,\pi^{2}\epsilon^3} \,\frac{(c_2)^2}{(b^2)^{-3\epsilon}}
\end{equation}
and
\begin{equation}\label{}
	\big(\tilde{\mathcal A}_2\big)_\mathrm{cl.}
	=
	\frac{(2m_Am_B\sigma)^5\sigma G^3}{2 \sqrt{\sigma^2-1}\,\pi^2\epsilon\, b^2}
	\left[
	4 c_2
	\left(
	c_0-\frac{3 c_2}{4m_A m_B\sigma}
	\right)
	+
	2c_2d
	\right].
\end{equation}
The leading eikonal $2\delta_0$ can be retrieved from eq.~\eqref{A0simple}, 
\begin{equation}\label{}
	2 \delta_0 = \frac{2 G m_A m_B \sigma^2 }{\sqrt{\sigma^2-1}}\frac{\Gamma(-\epsilon)}{(\pi  b^2)^{-\epsilon}}\,,
\end{equation}
while the leading quantum remainder $2\Delta_1$ is obtained from \eqref{A1simple} 
\begin{equation}\label{}
	2\Delta_1 = \frac{(4m_A m_B \sigma)^3 G^2\sigma}{4\sqrt{\sigma^2-1}} \frac{1}{\pi b^2}\,\left( c_0 - \frac{c_2}{m_A m_B \sigma} \right).
\end{equation}
The leading divergent part of $i\big(\tilde{\mathcal A}_2\big)_\mathrm{sscl.}$ cancels against that of $\frac{1}{3!}(2i\delta_0)^3$, as expected, while the divergent part of $2\delta_2$ is obtained subtracting $2\delta_0 \, 2i\Delta_1$ from $\big(\tilde{\mathcal A}_2\big)_\mathrm{cl.}$ according to the eikonal recipe discussed in Section~\ref{sec:eikonal}, so that
\begin{equation}\label{}
	2\delta_2 \simeq  \big(\tilde{\mathcal A}_2\big)_\mathrm{cl.}
	- 2\delta_0 \,2i\Delta_1
	=
	-\frac{i}{\pi \epsilon}\,\frac{16 m_A^2 m_B^2 \sigma^5 G^3}{b^2(\sigma^2-1)^2}\left[
	\sigma + (\sigma^2-2)\,\frac{\arccosh\sigma}{\sqrt{\sigma^2-1}}
	\right].
\end{equation}
This is purely imaginary and coincides with the one in \eqref{3.6}.

In fact, it is easy to obtain a more refined result showing that all divergences appearing in the ``doubly superclassical" part $i\big(\tilde{\mathcal A}_2\big)_\mathrm{sscl.}$ are consistent with the exponentiation of $2i\delta_0$, and not only the leading pole as checked above. 
To this end one needs to retain the $\epsilon$-exact expression for the one-loop amplitude to leading order in $q^2$. In this way eq.~\eqref{A1simple} takes the form 
\begin{equation}\label{}
		\mathcal A_1 = \frac{64(m_A m_B \sigma)^4G^2}{(4\pi)^{-\epsilon}}  \frac{C_2}{(q^2)^{1+\epsilon}}+\cdots\,,
\end{equation}
where $C_2$ is the contribution of box plus crossed box integrals to leading order in $q^2$ \cite{KoemansCollado:2019ggb,Cristofoli:2019neg}
\begin{equation}\label{}
	C_2 = \frac{\pi\epsilon}{\sin(\pi\epsilon)}\frac{\Gamma(-\epsilon)}{\Gamma(1-2\epsilon)}\frac{i \pi}{m_A m_B \sqrt{\sigma^2-1}}
\end{equation}
and reduces to $\tfrac1\epsilon\,c_2$ for small $\epsilon$.
Of course, this expression obeys the first eikonal exponentiation in impact-parameter space,
\begin{equation}\label{}
	i\tilde{\mathcal A_1} = \frac{(2i\delta_0)^2}{2!}+\cdots
\end{equation}
for any $\epsilon$.
Moreover, using once again the IR exponentiation in momentum space, according to which all IR poles are captured by \eqref{A2A1A0}, one obtains
\begin{equation}\label{}
	\big(\tilde{\mathcal A}_2\big)_\mathrm{sscl.}
	=
	\frac{(Gm_Am_B\sigma^2)^3}{\left(\sigma ^2-1\right)^{3/2}}
	\frac{4^{1+\epsilon} \pi ^{\frac{3}{2}+3\epsilon}   \cot (\pi  \epsilon )
		 \Gamma (-\epsilon )}{ \Gamma \left(\frac{1}{2}-\epsilon
		\right)}\frac{\Gamma(-3\epsilon)}{(b^2)^{-3\epsilon}}\,,
\end{equation}
and 
\begin{equation}\label{}
	i\big(\tilde{\mathcal A}_2\big)_\mathrm{sscl.}
	-
	\frac{(2i\delta_0)^3}{3!}
	=
	\frac{(Gm_Am_B\sigma^2)^3}{\left(\sigma ^2-1\right)^{3/2}}
	\,8 i \zeta_3
	+\mathcal O(\epsilon)
\end{equation}
showing that all $\mathcal O(\tfrac1{\epsilon^3})$, $\mathcal O(\tfrac1{\epsilon^2})$, $\mathcal O(\tfrac1{\epsilon})$ poles do cancel out, leaving behind only a finite mismatch (here $\zeta$  is the Riemann zeta function). 

\subsection{General Relativity}
\label{sec:GR}
The exponentiation of IR divergences in GR was discussed in Weinberg's celebrated paper \cite{Weinberg:1965nx} for arbitrary external particles. Rephrasing the original argument in the present language, in particular replacing the IR cutoff by dimensional regularization, one can isolate all infrared divergences of a given gravity amplitude $\mathcal A$ writing (see \eqref{loopintegral} for the precise definition of $\int_\Lambda d\ell$)
\begin{equation}\label{WeinbergResummation}
	\mathcal A = \mathcal A^0\, e^\mathcal{W}\,, 
	\qquad 
	\mathcal W = 
	\frac{1}{2(2\pi)^{4-2\epsilon}} \int_\Lambda d\ell\, B(\ell)\,,
\end{equation}
where $\mathcal A^0$ is IR finite and  $B(\ell)$ arises from a sum over all possible ways of attaching an IR graviton line to the external lines of the process,
\begin{equation}\label{Bgen}
	B(\ell)
	=
	-i\,\frac{8\pi G}{\ell^2-i0} \sum_{n,m} \frac{
		(p_n p_m)^2-\frac{1}{2-2\epsilon}\,m_m^2 m_n^2
		}{(\eta_np_n \ell-i0)(-\eta_m p_m\ell-i0)}
\end{equation}
with $\eta_n=+1$ if $n$ is an outgoing line and $\eta_m=-1$ for an incoming line. The leftover integrals in $\mathcal W$ can be evaluated using \eqref{general}, \eqref{generalI} and \eqref{special}, and are of course independent of the small cutoff $\Lambda$ to leading order in $\epsilon$.

To this order, specializing \eqref{WeinbergResummation} to the amplitude of interest, one obtains,	
\begin{align}\label{WeinbergW}
	\mathcal W &= \frac{G}{2\pi\epsilon} \sum_{n,m=1}^{4}w_{nm}\,,
	\\
	w_{nm} &= m_n m_m \left(\sigma_{nm}^2-\frac{1}{2}\right)
	\frac{\eta_{n}\eta_{m} \operatorname{arccosh}\sigma_{nm}-i\pi \theta_{nm}}{\sqrt{\sigma_{nm}^2-1}}\,,
\end{align}
where
\begin{align}\label{key}
	p_1&=p_A\,,\quad p_2=p_B\,,\quad
	p_3=p'_B\,,\quad p_4=p'_A\,,\\
	m_1&=m_A\,,\quad m_2=m_B\,,\quad
	m_3=m_B\,,\quad m_4=m_A\,,\\
	\eta_{1}&=\eta_2=-1\,, \quad \eta_3=\eta_4=+1\,,\quad
	\sigma_{nm}=-\frac{p_n p_m}{m_n m_m}
\end{align}  and
\begin{equation}\label{}
	\theta_{nm}=\begin{cases}
		1 \quad &\text{if } (n,m)\in\{(1,2),(2,1),(3,4),(4,3)\}\\
		0 \quad &\text{otherwise}\,.
	\end{cases}
\end{equation}
In particular, the right prescription to avoid the pinching singularities arising for $n=m$ is to simply drop the formally infinite imaginary pieces \cite{Yennie:1961ad} writing $w_{11}=w_{44}=\tfrac12 m_A^2$ and $w_{22}=w_{33}=\tfrac12m_B^2$. Let us also note that
\begin{equation}\label{}
	\sigma_{12} = \sigma\,,\qquad
	\sigma_{13} = \sigma - \frac{q^2}{2m_A m_B}\,, \qquad
	\sigma_{14} = 1 + \frac{q^2}{2m_A^2}\,, \qquad 
	\sigma_{23} = 1 + \frac{q^2}{2m_B^2}\,,
\end{equation}
and similarly for the remaining terms.
Therefore, although divergent as $\epsilon\to0$, the exponent $\mathcal W$ is analytic in $q^2$.

From the explicit loop expansion
\begin{equation}\label{}
	\mathcal A = \mathcal A_0+\mathcal A_1+ \mathcal A_2+\cdots= e^{\mathcal W} (\mathcal A_0+\mathcal A^0_1+ \mathcal A^0_2+\cdots)\,,
\end{equation}
one obtains
\begin{align}\label{}
	\label{exp1}
	\mathcal A_1 &= \mathcal A_1^0 + \mathcal W \mathcal A_0\,,\\
	\label{exp2}
	\mathcal A_2 &= \mathcal A_2^0 + \mathcal W \mathcal A_1^0 + \frac12 \mathcal W^2\mathcal A_0^2\,,
\end{align}
and therefore the IR-divergent part of the two-loop amplitude $\mathcal A_2$ is given by $\mathcal W \mathcal A_1 -\frac{1}{2}\mathcal W^2\mathcal A_0$. Moreover, the tree-level exchange is given by \cite{KoemansCollado:2019ggb,Cristofoli:2020uzm} 
\begin{equation}\label{treeLGR}
	\mathcal A_0 = 2(8 \pi G)\,\frac{\gamma_s}{q^2}\,,
	\qquad
	\gamma_s = 2m_A^2 m_B^2\,\left(\sigma^2-\tfrac{1}{2-2\epsilon}\right),
\end{equation}
up to analytic terms, so that the sought-after $\frac{1}{\epsilon}\log q^2$ terms in $\mathcal A_2$ must arise from terms within $\mathcal W \mathcal A_1$ proportional to $\frac{1}{\epsilon^2(q^2)^{\epsilon}}$, which we denote as above
\begin{equation}\label{homq}
	\big(
	\mathcal A_2
	\big)_\mathrm{cl.}
	=
	\left[
	\mathcal W \mathcal A_1
	\right]_{\mathcal O\big(\frac{1}{\epsilon^2(q^2)^{\epsilon}}\big)}.
\end{equation}
In order to obtain these terms, it is enough to retain the divergent contributions to the one-loop amplitude that stem from box and crossed box topologies \cite{KoemansCollado:2019ggb,DiVecchia:2021bdo}, accurate up to $\mathcal O\left((q^2)^{-\epsilon}\right)$,
\begin{equation}\label{A1grav}
	\mathcal A_1=\frac{16 G^2}{\epsilon(q^2)^{\epsilon}} \left[\gamma_s^2 \frac{c_{\mathrm{II}}}{q^2}+\gamma_u^2\left(\frac{c_{\mathrm{X}}}{q^2}+c_{0}\right)+\cdots\right]
\end{equation}
with
\begin{equation}\label{}
	\gamma_u \simeq \gamma_s - 2 m_A m_B \sigma q^2 \,.
\end{equation}
The reason for this simplification is that focusing on non-analytic contributions allows one to neglect a number of topologies, such as bubbles or self-energy diagrams, that may only give rise to contact divergences. Moreover, triangle topologies do give rise to non-analytic terms, but not to divergent contributions in $\epsilon$.
Proceeding in this way, using in particular eq.~\eqref{homq}, one is led to the two-loop divergent part
\begin{equation}\label{WA1}
\big(
\mathcal A_2
\big)_\mathrm{cl.}
	\simeq
	- \frac{i (4m_A m_BG)^3}{2\epsilon^2(q^2)^{\epsilon}(\sigma^2-1)^{\frac32}}
	\left(\sigma^2-\tfrac12\right)^2\left[
	\frac{\sigma^2+5}{3}
	+ 2\sigma(4\sigma^2-5)\,\frac{\arccosh\sigma}{\sqrt{\sigma^2-1}}
	\right].
\end{equation}	
Let us also mention that, combining \eqref{WeinbergW}, \eqref{treeLGR} and \eqref{A1grav}, the non-analytic IR divergence of the type $\epsilon^{-1}(q^2)^{-1}$ indeed cancels out in the combination $\mathcal A_1-\mathcal W \mathcal A_0$, in accordance with \eqref{exp1}. In order to check the cancellation of the analytic IR divergence proportional to $\epsilon^{-1}(q^2)^{0}$, one ought to include contributions to \eqref{A1grav} analogous to the $d$-term in eq.~\eqref{A1simple}. However, being analytic, these contributions are not needed in order to calculate the IR-divergent part of $2\delta_2$ via \eqref{exp2} and can be safely discarded for the present purposes, as explained above.

The tree-level exchange \eqref{treeLGR} determines the leading eikonal 
\begin{equation}\label{}
	2\delta_0  = \frac{2Gm_A m_B}{\sqrt{\sigma^2-1}}\left(\sigma^2-\tfrac{1}{2-2\epsilon}\right)\frac{\Gamma(-\epsilon)}{(\pi b^2)^{-\epsilon}}
\end{equation}
after going to impact-parameter space with the help of eqs.~\eqref{2dFT}, \eqref{basic}.
The leading remainder $2\Delta_1$ can be obtained from \eqref{A1grav}, to leading order in $\epsilon$,
\begin{equation}\label{}
	2\Delta_1 = \frac{4 G^2}{\pi m_A m_B\sqrt{\sigma^2-1} b^2} \,\gamma_s (\gamma_s c_0-4 m_A m_B \sigma c_\mathrm{X})\,.
\end{equation} 
To calculate the divergent contribution to $2\delta_2$, it is then sufficient to translate \eqref{WA1} to impact-parameter space and subtract $2\delta_0\,2i\Delta_1$, 
\begin{equation}\label{}
		2\delta_2 \simeq 
	\big(
	\tilde{\mathcal A}_2
	\big)_\mathrm{cl.}
	-
	2\delta_0\,2i\Delta_1\,.
\end{equation}
This yields 
\begin{equation}\label{}
		2\delta_2 \simeq 
	-\frac{i}{\pi\epsilon}
	\frac{(2G)^3m_A^2 m_B^2\left(
		\sigma^2-\tfrac12
		\right)^2}{b^2(\sigma^2-1)^2}
	\left[
	\frac{8-5\sigma^2}{3}
	+
	\sigma
	(2\sigma^2-3)\,
	\frac{\arccosh\sigma}{\sqrt{\sigma^2-1}}
	\right].
\end{equation}
Again this is a purely imaginary divergent part which agrees with the one in eq.~\eqref{3.2GR}.

\subsection*{Acknowledgments}
I would like to thank Henrik Johansson and Gabriele Veneziano for useful comments. I am also grateful to Paolo Di Vecchia for carefully going through a preliminary version of this manuscript, to Rodolfo Russo for very insightful discussions on IR divergences in maximal supergravity, and to Stephen Naculich for stimulating observations that led to an improved understanding of several points.
This work is supported by the Knut and Alice
Wallenberg Foundation under grant KAW 2018.0116.

\appendix

\section{Notation and Conventions}
\label{app:conventions}
The mostly-plus metric is adopted, so that space-like (time-like) vectors square to positive (negative) values. For $d$-dimensional integrals and delta functions, the shorthand expressions $\hat dq$ and $\hat\delta(q)$ stand for
\begin{equation}\label{}
	\hat dq = \frac{dq}{(2\pi)^d}\,,\qquad \hat \delta(q) = (2\pi)^d\delta(q)\,.
\end{equation}
The integrals $\int_\ell$ and $\int_\Lambda d\ell$ are defined as follows
\begin{equation}\label{loopintegral}
	\int_\ell = \frac{1}{i\pi^{2-\epsilon}}\int d\ell\,,
	\qquad
	\int_\Lambda d\ell = \int_{|\vec \ell\,|<\Lambda} \!\!\! d\vec\ell \int_{-\infty}^{+\infty} d\ell^0\,,
\end{equation}
where $\ell$ is $(4-2\epsilon)$-dimensional, while $\ell=(\ell^0,\vec\ell\,)$ with $\vec\ell$ a $(3-2\epsilon)$-dimensional vector.

The transverse Fourier transform is defined by
\begin{equation}\label{IPFT}
	\tilde  f(b) = \int 
	\hat \delta(2p_Aq) \, 
	\hat \delta(2p_Bq)\,  e^{ibq} f(q) \, \hat dq\,,
\end{equation}
and it can be recast in form
\begin{equation}\label{IPFTcm}
	\tilde f(b) =\frac{1}{4\Omega}  \int e^{ibq_\perp} f(q_\perp) \hat dq_\perp\,,
	\qquad
	\Omega=  m_A m_B \sqrt{\sigma^2-1}\,,
\end{equation}
where $q_\perp$ lies in the transverse $(2-2\epsilon)$-dimensional space defined by $p_Aq_\perp=0=p_Bq_\perp$. 
Fourier transforms of this type can be calculated via
\begin{equation}\label{basic}
	\int \frac{e^{i b q_\perp}}{(q_\perp^2)^\alpha} \,\hat dq_\perp = \frac{1}{\pi^{1-\epsilon}4^\alpha \Gamma(\alpha)}\,\frac{\Gamma(1-\epsilon-\alpha)}{(b^2)^{1-\epsilon-\alpha}}\,.
\end{equation}

For completeness, let us include a quick derivation of eq.~\eqref{basic}, starting from
\begin{equation}\label{transff}
	\hat f(b) = \int e^{ibq}\,f(q)\, \hat dq \,,\qquad
	f(q)=\frac1{(q^2)^{\alpha}}
\end{equation}
in $d$ Euclidean dimensions. Since $\hat f(Rb) = \hat f(b)$ for any $d$-dimensional rotation $R$ and $\hat f(\lambda b) = \lambda^{2\alpha-d}\hat f(b)$ for any scale $\lambda$, one sees that $\hat f(b)$ must take the form
\begin{equation}\label{transffhat}
	\hat f(b) = \frac{C}{(b^2)^{\frac{d}{2}-\alpha}}\,,
\end{equation}
for some constant $C$.
On the other hand, denoting Fourier transforms with a hat as in \eqref{transff}, 
\begin{equation}\label{fgidentity}
	\int f(q)\, \hat g(q) \,dq = \int \hat f(b) \, g(b) \,db\,,
\end{equation}
for any $f$, $g$.
Choosing $f$, $\hat f$ as in \eqref{transff}, \eqref{transffhat}, and 
\begin{equation}\label{}
	\hat g(b) =  e^{-\frac{b^2}{2}}\,,
	\qquad 
	g(q) = (2\pi)^{\frac{d}{2}}\,e^{-\frac{q^2}{2}} \,,
\end{equation}
the condition \eqref{fgidentity} translates into
\begin{equation}\label{}
\int \frac{e^{-\frac{q^2}{2}}}{(q^2)^{\alpha}} \, dq 
=
C\,
(2\pi)^{\frac{d}{2}}
\int \frac{e^{-\frac{b^2}{2}}}{(b^2)^{\frac{d}{2}-\alpha}} \, db\,.
\end{equation}
The remaining integrals can be evaluated in terms of Gamma functions, leading to
\begin{equation}\label{}
	C = \frac{\Gamma\left(\frac{d}{2}-\alpha\right)}{\pi^{\frac{d}{2}}4^{\alpha}\Gamma(\alpha)}\,,
\end{equation}
so that in conclusion
\begin{equation}\label{}
	\int
	\frac{e^{iqx}}{(q^2)^{\alpha}}\,\hat dq
	=\frac{1}{\pi^{\frac{d}{2}}4^{\alpha}\Gamma(\alpha)}\,\frac{\Gamma\left(\frac{d}{2}-\alpha\right)}{(b^2)^{\frac{d}{2}-\alpha}}\,,
\end{equation}
which reduces to \eqref{basic} for $d=2-2\epsilon$.

\section{IR Integrals}
\label{app:IR}
The goal of this appendix is to evaluate the leading IR-divergent part of the integrals
\begin{equation}\label{defIpm}
	I_\mp(p,p') = \int_\Lambda\frac{d\ell}{(2p\ell\mp i0)(2p'\ell - i0)(\ell^2-i0)}
\end{equation}
for two future-directed time-like vectors $p$, $p'$. Going to a reference frame where $p=(E,\vec p\,)$, $p'=(E',-\vec p\,)$ yields
\begin{equation}\label{Iint}
	I_\mp= \int_{|\vec\ell\,|<\Lambda}\!\!\! d\vec\ell\int_{-\infty}^{+\infty}\frac{d\ell^0}{(-2E\ell^0+2\vec p \,\vec \ell\mp i0)(-2E'\ell^0-2\vec p \, \vec \ell-i0)(-(\ell^0)^2+|\vec\ell\,|^2-i0)}\,.
\end{equation}
The integral over $\ell^0$ can be evaluated via residues. Focusing first on $I_-$ and closing the contour in the upper half-plane leads to
\begin{equation}\label{}
	I_-= i\pi\int_{|\vec\ell\,|<\Lambda}\frac{d\vec\ell}{(2E|\vec\ell\,|+2\vec p \, \vec \ell\,)(2E'|\vec\ell\,|-2\vec p\,\vec\ell\,)|\vec\ell\,|}\,, 
\end{equation}
and, going to polar coordinates $\vec \ell = \omega \hat n$ with $|\hat n|=1$,
\begin{equation}\label{Jint}
	I_-= \frac{i\pi}{4} \int_0^\Lambda \frac{d\omega}{\omega^{1+2\epsilon}}\int\frac{d\Omega(\hat n)}{(E+\vec p \, \hat n)(E'-\vec p\,\hat n)}\,.
\end{equation}
For small $\epsilon$ with a negative real part,
\begin{equation}\label{}
	\int_0^\Lambda \frac{d\omega}{\omega^{1+2\epsilon}}
	=
	\left[-\frac{\omega^{-2\epsilon}}{2\epsilon}\right]_0^\Lambda = -\frac{\Lambda^{-2\epsilon}}{2\epsilon} = -\frac{1}{2\epsilon} + \mathcal O(\epsilon^0)\,, 
\end{equation}
so that \eqref{Jint} can be estimated  as follows to leading order in $\epsilon$, evaluating the two-dimensional angular integral
\begin{equation}\label{}
	I_-=- \frac{i\pi^2}{4\epsilon}\int_{-1}^{+1}\frac{dx}{(E+|\vec p\,|x)(E'-|\vec p\,|x)}
	=\frac{-i\pi^2}{4\epsilon(E+E')|\vec p\,|}\,\log\frac{(E+|\vec p\,|)(E'+|\vec p\,|)}{(E-|\vec p\,|)(E'-|\vec p\,|)}\,.
\end{equation}
This result can be cast in the invariant form
\begin{equation}\label{}
	I_-(p,p') = K(p,p') \equiv -\frac{1}{\epsilon} \frac{i\pi^2}{2mm'}\, \frac{\arccosh\sigma_{pp'}}{\sqrt{\sigma^2_{pp'}-1}}\,,
\end{equation}
where 
\begin{equation}\label{symbols}
	\sigma_{pp'} = -\frac{p\,p'}{mm'}\,,\qquad   p^2=-m^2\,,\qquad p'^2=-m'^2
\end{equation}
with positive $m$, $m'$. 

In the calculation of $I_+$, the integral in \eqref{Iint} receives an additional contribution from the pole at $\ell^0=\vec p\,\vec\ell/E+i0$,
\begin{equation}\label{}
	I_+(p,p')=K(p,p')+\frac{i\pi}{2}\int_{|\vec\ell\,|<\Lambda}\frac{d\vec\ell}{(E+E')\big(\vec p\,\vec \ell+i0\big)\Big[-\tfrac1{E^2}\big(\vec p\,\vec \ell\ \big)^2+|\vec\ell\,|^2\Big]}
\end{equation}
and performing the $\omega$ integral as above,
\begin{equation}\label{}
	I_+=K-\frac{i\pi^2}{2\epsilon}\int_{-1}^{+1}\frac{dx}{(E+E')|\vec p\,|\big(x+i0\big)(-\tfrac{|\vec p\,|^2}{E^2}\,x^2+1)}\,.
\end{equation}
Using $\frac1{x+i0}=\mathrm{PV}\frac1{x}-i\pi\delta(x)$ and noting that, by parity, only $-i\pi\delta(x)$ can contribute, one is thus left with
\begin{equation}\label{}
	I_+
	=K-\frac{\pi^3}{2\epsilon}\frac{1}{(E+E')|\vec p\,|}\,.
\end{equation}
The result can be cast in the invariant form
\begin{equation}\label{}
	I_+(p,p')= -\frac{1}{\epsilon} \frac{i\pi^2}{2mm'}\, \frac{\arccosh\sigma_{pp'}-i\pi}{\sqrt{\sigma^2_{pp'}-1}}\,.
\end{equation}

In conclusion, in the notation of eq.~\eqref{symbols},
\begin{align}\label{general}
\int_{\Lambda}\frac{d\ell}{(2p\ell - i0)(2p'\ell - i0)(\ell^2-i0)}
&=-\frac{1}{\epsilon} \frac{i\pi^2}{2mm'}\, \frac{\arccosh\sigma_{pp'}}{\sqrt{\sigma^2_{pp'}-1}}\,,\\
\label{generalI}
\int_{\Lambda}\frac{d\ell}{(2p\ell + i0)(2p'\ell - i0)(\ell^2-i0)}
&=-\frac{1}{\epsilon} \frac{i\pi^2}{2mm'}\, \frac{\arccosh\sigma_{pp'}-i\pi}{\sqrt{\sigma^2_{pp'}-1}}\,.
\end{align}
Note that $I_+(p,p')=I_+(p',p)$ and $I_-(p,p')=I_-(p',p)$, so that \eqref{general} and \eqref{generalI} exhaust the only two independent sign choices for the linearized propagators. Moreover, $I_-(p,p)$ is simply obtained by sending $\sigma_{pp'}\to1$ and setting $m=m'$ in \eqref{general},
\begin{equation}\label{}
	\label{special}
	\int_{\Lambda}\frac{d\ell}{(2p\ell - i0)^2(\ell^2-i0)}
=-\frac{1}{\epsilon} \frac{i\pi^2}{2m^2}\,,
\end{equation}
as can be seen going to a frame where $p=(m,0)$ and retracing the above steps.

\providecommand{\href}[2]{#2}\begingroup\raggedright\endgroup

\end{document}